\documentclass[letterpaper]{article}

\usepackage{natbib,alifeconf}  

\usepackage[hidelinks]{hyperref}
\usepackage{cleveref}
\usepackage[small]{caption}
\usepackage[font=scriptsize,labelfont=scriptsize]{subcaption}
\usepackage{url}

\title{On the Existence of Information Bottlenecks in \\Living and Non-Living Systems}
\author{Michael Crosscombe \and Hiroki Sato\\
\mbox{}\\
Graduate School of Arts and Sciences, University of Tokyo, Japan \\
cross@sacral.c.u-tokyo.ac.jp} 

\begin{document}
\maketitle

\section{Introduction}

Within the field of Artificial Life (ALife), a lot of interest has built up around different ways to frame various topics of ongoing research; ways that enable us to adopt new perspectives from which to think about and to tackle the open problems in the field.
A particularly challenging problem in the field is that of emergence. Specifically, emergent behaviour is the spontaneous existence of new and unexpected patterns, structures, or behaviours that arise from the interactions of simpler components in a system. For example, in social insect colonies such as ants and honey bees, the observable collective behaviours emerge from the interactions between the individual insects, rather than some top-down coordination.
In collective systems more generally, individuals are forced to develop novel ways of compressing or encoding information that can be propagated efficiently to other agents, before it is decompressed or decoded, leading to the emergence of intelligent collective behaviours.

In many complex systems, we observe that `interesting behaviour' is often the consequence of a system exploiting the existence of an \emph{Information Bottleneck} (IB).
These bottlenecks can occur at different scales, between individuals or components of a system, and sometimes within individuals themselves.
Oftentimes, we regard these bottlenecks negatively; as merely the limitations of the individual's physiology and something that ought to be overcome when designing and implementing artificial systems.
However, we suggest instead that IBs may serve a purpose beyond merely providing a minimally-viable channel for coordination in collective systems.
More specifically, we suggest that \emph{interesting} or \emph{novel} behaviour occurs when the individuals in a system are constrained or limited in their ability to share information and must discover novel ways to exploit existing mechanisms, which are inherently bottlenecked, rather than circumventing or otherwise avoiding those mechanisms entirely.



\section{Information is Everything}

There has been a strong interest of late to approach the world through the lens of \emph{information}. For example, \citet{Gershenson2012-info} proposed to view the fundamental units of life and matter as information, rather than energy. Many ideas surrounding this central theme, such as how information emerged, the role of noise in living systems, and state space compression have been explored in great detail by \citet{Walker2017}. More recently, \citet{Gershenson2023-emergence} has attempted to frame emergent phenomena across a broad spectrum of ALife systems in terms of information and its properties in living (biological) and non-living (artificial) systems.
Another perspective in active development by \citet{Flack2017-CG} and colleagues is that of coarse-graining, which adopts a similar information-based view but sees emergence as a process in which information is aggregated and simplified as it moves from the micro scale to the macro scale, making available a more stable, accurate aggregate representation to the micro-scale components as a result.



In the seminal paper ``Driven by compression progress'', \citet{Schmidhuber2008-compression} described how curiosity -- the driving force of all artistic and scientific endeavours -- can be framed as the desire to discover novel and surprising representations of information. When previously unknown regularities are discovered, new ways to compress information lead to progress. Yet there remains some utility in systems exploiting simple, efficient and perhaps even \emph{noisy} means of communication to achieve coordination, as discussed recently by \citet{Kahneman2022}.
All of these related endeavours within the field point to the need for a system to gather, process, and propagate information in an efficient manner, and we propose that IBs may present a valuable perspective that can be observed in nature and adopted in the implementation of artificial systems to attain emergent collective behaviours.

\section{Examples of Information Bottlenecks}

\paragraph{Genetic bottleneck.}
Let's first consider the human brain.
This highly complex system consists of about $86$ billion neurons, each with an average of around $7\,000$ synaptic connections~\citep{azevedo2009equal}. However, the human genome measures about $3$ billion basepairs in length~\citep{HumanGenome}.
The human genome can be encoded in approx.\ $\mathbf{750}$ \textbf{megabytes}, but encoding the entire connectivity of the human brain would require approx.\ $\textbf{4.6}$ \textbf{petabytes}\footnote{Due to a lack of space, we refer readers to the calculations of \citet{Zador2019critique}, though we have adopted different reference values.}.
Of course, this is only true if we were to assume that the \emph{entirety} of the human genome was dedicated to encoding the structures of the brain, which we know cannot be the case. This gives way to the concept of the \emph{genomic bottleneck}, or the idea that the genome cannot explicitly specify every connection in the brain and, therefore, must instead provide a ruleset by which the brain becomes a self-organising structure attuned to rapid learning. That is, it can be said that the brain's structure emerges from the genome via an information bottleneck. 

\paragraph{Pheromone trails.}
The use of pheromones by ants is perhaps the most studied example of stigmergy~\citep{Grasse1959}, or the process by which a collective system can achieve some level of self-organisation through modification of its environment~\citep{Theraulaz1999}.
Individual ants use trail pheromones~\citep{Wilson1962} to produce stable structures that facilitate coordinated behaviour between individuals.
In this way, pheromone trails are an example of a coordination mechanism which facilitates collective behaviour via a bottleneck \emph{between individuals} in the colony.

\begin{figure}[t]
    \centering
    \includegraphics[width=0.47\textwidth]{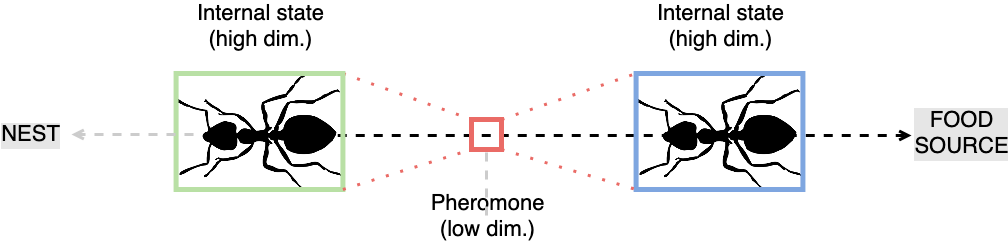}
    \caption{An illustration of the information bottleneck in ants. An ant's discovery of a food source triggers a secretion of the trail pheromone. This can be considered an encoding of the high-dimensional information of the internal state to the low-dimensional information via the information bottleneck that is the ant's physiology.}
    \label{fig:IB}
\end{figure}


We attempt to depict this in \Cref{fig:IB}.
From this minimal form of information, other ants are able to extrapolate the placement of pheromones in relation to the nest, with additional contextual information, to locate the food source.

\paragraph{Non-living systems}
Recently, there has been increasing interest in the presence of constrained forms of information sharing -- of IBs -- in many bio-inspired models of collective behaviours. In swarm robotics, for example,
\citet{Talamali2021} and \citet{Aust2022} have studied the degradation in performance as these bottlenecks are widened, meaning the communication constraints in the system were lifted.
Similar studies consider the impact of interaction constraints in autonomous systems which communicate via explicitly predefined interaction networks~\citep{Crosscombe2021,Crosscombe2022,Kwa2023}.
Generally, as information sharing in these systems becomes less constrained, performance decreases. As bottlenecks in living systems often result from the physiology, we then presume that we can greatly improve performance by widening these bottlenecks and lifting the constraints in artificial systems. Yet it would appear that doing so often worsens performance, hinting at the importance of constrained mechanisms for emergence to occur.

\subsection{The role of the Information Bottleneck}


Broadly, in a collective system in which the individuals are involved in the processing of information in pursuit of a shared goal, an individual must \emph{encode} some internally-held information and then propagate this information in a way that allows them to communicate something (about their own state, or the state of their environment) to others in the system. Other individuals in the system must, in turn, \emph{decode} and interpret the encoded information, perhaps in conjunction with additional context. In general, this describes many systems of collective behaviour, including ants and honey bees~\citep{Biesmeijer2005-WaggleDance}, but also many ALife systems such as the Boid model~\citep{Reynolds1987}, and most recently the work of \citet{Fox2023} in which they impose a bottleneck in pursuit of emergent communication.
We argue that IBs may play a fundamental role, particularly in collective systems, by driving the system to explore novel ways of compressing or encoding information that can be utilised by the individual components of the system. As the processes of encoding/decoding are not \emph{lossless} (in compression terms) they result in variation in the reconstructed information. We believe this is beneficial because noisy interpretations of information can increase exploration, resulting in the discovery of more optimal ways to utilise existing mechanisms for collective coordination.


\section{Future Experiments}

To explore this concept, we propose to simulate the evolution of a series of collective behaviours through varying degrees of information bottlenecks. Beginning with simple behaviours, such as flocking, and progressing to more complex behaviours, we can explore when individuals are able to exploit existing information pathways for effective coordination and when those pathways are too (un)restricted to lead to the emergence of collective behaviours.
In parallel, we also intend to conduct experiments using colonies of \textit{Pristomyrmex punctatus} ants~\citep{Tsuji2011social} to determine what levels of density and connectivity are sufficient, or necessary, for the emergence of collective behaviours observed in this species (e.g., clustering).

\section{Acknowledgements}

We thank the reviewers for their valuable suggestions.
Michael Crosscombe is a Japan Society for the Promotion of Science (JSPS) international research fellow hosted by Prof.\ Takashi Ikegami.

\footnotesize
\bibliographystyle{apalike}
\bibliography{references} 

\end{document}